\documentclass[conference]{IEEEtran}
\usepackage[T1]{fontenc}

\ifCLASSINFOpdf
\else
\fi

\usepackage{subfigure}
\usepackage{float}
\usepackage{boxedminipage}
\usepackage{graphicx}
\usepackage{amsmath}
\usepackage{amssymb}
\usepackage{tikz}
\newcommand{\twopartdef}[4]
{
	\left\{
	\begin{array}{ll}
		#1 & \mbox{if } #2 \\
		#3 & \mbox{otherwise,} #4
	\end{array}
	\right.
}

\begin{document}

\title{Cooperative Slotted ALOHA for Massive M2M Random Access Using Directional Antennas}

\author{\IEEEauthorblockN{Aleksandar Mastilovic, Dejan Vukobratovic}
\IEEEauthorblockA{Department of Power, Electronics\\and Communications Engineering,\\
University of Novi Sad, Serbia\\
Email: \{aleksandar.mastilovic,dejanv\}@uns.ac.rs}
\and
\IEEEauthorblockN{Dusan Jakovetic, Dragana Bajovic}
\IEEEauthorblockA{BioSense Institute\\
University of Novi Sad, Serbia\\
Email: \{djakovet,dbajovic\}@uns.ac.rs }
}

\maketitle

\begin{abstract}
Slotted ALOHA (SA) algorithms with Successive Interference Cancellation (SIC) decoding have received significant attention lately due to their ability to dramatically increase the throughput of traditional SA. Motivated by increased density of cellular radio access networks due to the introduction of small cells, and dramatic increase of user density in Machine-to-Machine (M2M) communications, SA algorithms with SIC operating cooperatively in multi base station (BS) scenario are recently considered. In this paper, we generalize our previous work on Slotted ALOHA with multiple-BS (SA-MBS) by considering users that use directional antennas. In particular, we focus on a simple randomized beamforming strategy where, for every packet transmission, a user orients its main beam in a randomly selected direction. We are interested in the total achievable system throughput for two decoding scenarios: i) non-cooperative scenario in which traditional SA operates at each BS independently, and ii) cooperative SA-MBS in which centralized SIC-based decoding is applied over all received user signals. For both scenarios, we provide upper system throughput limits and compare them against the simulation results. Finally, we discuss the system performance as a function of simple directional antenna model parameters applied in this paper.       
\end{abstract}

\begin{IEEEkeywords}
Slotted ALOHA, successive interference cancellation, M2M, directional antennas
\end{IEEEkeywords}

\IEEEpeerreviewmaketitle

\section{Introduction}

The problem of connecting very large number of devices to wireless cellular networks is gaining momentum as billions of devices are estimated to be connected to the Internet as part of so called M2M communications. In mobile cellular networks, the first M2M standards have been introduced under the name Machine Type Communication (MTC) services as part of the fourth generation (4G) 3GPP Long-Term Evolution (LTE) technology and will continue to evolve into the upcoming fifth generation (5G) technology \cite{MTC}. The vast majority of MTC users will be devices whose activity is irregular and unpredictable and that occasionally transmit small volumes of data such as smart meters. As the cellular infrastructure becomes increasingly dense due to proliferation of small cells, the 5G radio access network is faced with increased density both in terms of user and base stations. In such a scenario, we are interested in the design of simple and efficient random access solution able to support the expected surge of MTC traffic in the future.

Slotted ALOHA (SA) random access solutions with Successive Interference Cancellation (SIC) decoding have received significant attention lately due to their ability to dramatically increase the throughput of traditional SA. SA with SIC for single base station systems has been proposed in \cite{CRDSA}. Using the analogy with sparse-graph codes and iterative erasure decoding, SA with SIC is further optimized to reach close-to-optimal throughputs \cite{IRSA}. Motivated by increased density of cellular networks due to the introduction of small cells, we recently considered SA algorithms with SIC operating cooperatively in multi base station (SA-MBS) systems \cite{SA-MBS}. In SA-MBS, users can be heard and decoded by any of the surrounding base stations as, from the system perspective, it is not important which of the small base stations collected the user. Thus, apart from temporal diversity exploited by SA with SIC in single base station systems, SA-MBS may additionally exploit spatial diversity combined with cooperative SIC-based decoding. For an overview of SA with SIC in single and multi base station systems, we point the reader to \cite{CSA} and \cite{MBS}.

In this paper, we generalize the work in \cite{SA-MBS} by considering users that use directional antennas. Our main motivation follows recent trends of shifting the operational band of mobile cellular systems towards millimetre wavelength (mmWave) domain and usage of directional transmissions by exploiting multiple antenna systems and beamforming techniques \cite{mmW-5G}\cite{mmW-BF}. In particular, we consider a simple randomized beamforming strategy where, for every packet transmission, a user orients its main beam in a randomly selected direction \cite{RBF}. This way, we maintain the simplicity of random access SA-MBS scheme, avoid complex and time-consuming beamforming alignment procedures, and rely on density of small cell infrastructure and efficiency of SA-MBS with joint SIC-based decoding. We are interested in the achievable system throughput for two decoding scenarios: i) non-cooperative scenario in which traditional SA operates at each BS independently, and ii) cooperative SA-MBS in which centralized SIC-based decoding is performed over all received user signals. The latter scenario is motivated by the case where  user signals are centrally processed as part of emerging Cloud RAN (C-RAN) architectures \cite{5G-Disrupt}. We provide analytical approximations for the system throughput and compare them against the results obtained via simulation experiments. The analysis is concluded by the discussion of the system throughput performance as a function of directional antenna model parameters such as the beam-width and the beam-range.

The rest of the paper is organized as follows. In Section 2, we present details of the SA-MBS system model with directional antennas. Section 3 introduces the two versions of the SA-MBS decoding and provides simple but useful approximations   of the system throughput. In Section 4, we present simulation results and compare them against the analytical approximations. Finally, Section 5 concludes the paper.

\section{System Model}

\subsection{Placement Model}

We assume that both BSs and UEs are placed according to Poisson 
point processes (PPP) over a surface $\mathcal{A}$ of an area $\lVert \mathcal{A} \rVert$. The PPP 
for BSs has intensity $\lambda_{BS}$, while for UEs it has intensity $\lambda_{UE}$. The two PPP 
are mutually independent. The numbers of BSs and UEs, denoted as $N_{BS}$ and $N_{UE}$, are hence 
random variables with Poisson distributions $\mathcal{P}(\overline{N}_{BS})$ and $\mathcal{P}(\overline{N}_{UE})$, with mean values $\overline{N}_{BS}=\lambda_{BS} \cdot \lVert \mathcal{A} \rVert$ and $\overline{N}_{UE}=\lambda_{UE} \cdot \lVert \mathcal{A} \rVert$, respectively. 
We denote users by $U_{i}$, $i=1,2,...,$ and BSs by
 $B_{j}$, $j=1,2,...$ Unless otherwise stated, we will focus on a unit-square area $\mathcal{A}$ ($\lVert \mathcal{A} \rVert=1$), in which case the expected number of BSs and UEs reduces to $\overline{N}_{BS}=\lambda_{BS}$ and $\overline{N}_{UE}=\lambda_{UE}$, respectively.

\subsection{Random Access Model}

We consider \textit{Slotted ALOHA} random access model in the \textit{Multi Base Station} (SA-MBS) scenario \cite{SA-MBS}. The time domain is discrete and divided into time slots (TS). User transmissions are synchronized and aligned with TS boundaries, which are perfectly synchronized across all BSs. In any time slot, a UE transmits an equal-length data packet independently of other UEs with probability $p$, which we call the activity factor. Due to high BS density, the UE packet transmission may be detected at several neighbouring BSs. We assume BSs are interconnected via a backhaul network and any BS may collect any UE's data packet, i.e., we assume no a priori UE to BS associations. We consider the UE's packet to be collected as long as it is decoded by any BS. 
 
 The average normalized load is defined by ${G}=p\frac{\overline{N}_{UE}}{\overline{N}_{BS}}=p\frac{\lambda_{UE}}{\lambda_{BS}}$.
 For the sake of analysis, without loss of generality, it is sufficient to consider the SA-MBS system behaviour at any single fixed TS. In the following, we will assume activity factor $p=1$. This is sufficient, as any other $p<1$ will only thin the PPP describing UE placement to intensity $p\lambda_{UE}$.

\subsection{User Transmission Model}

\begin{figure}
\centering
\includegraphics[width=3in]{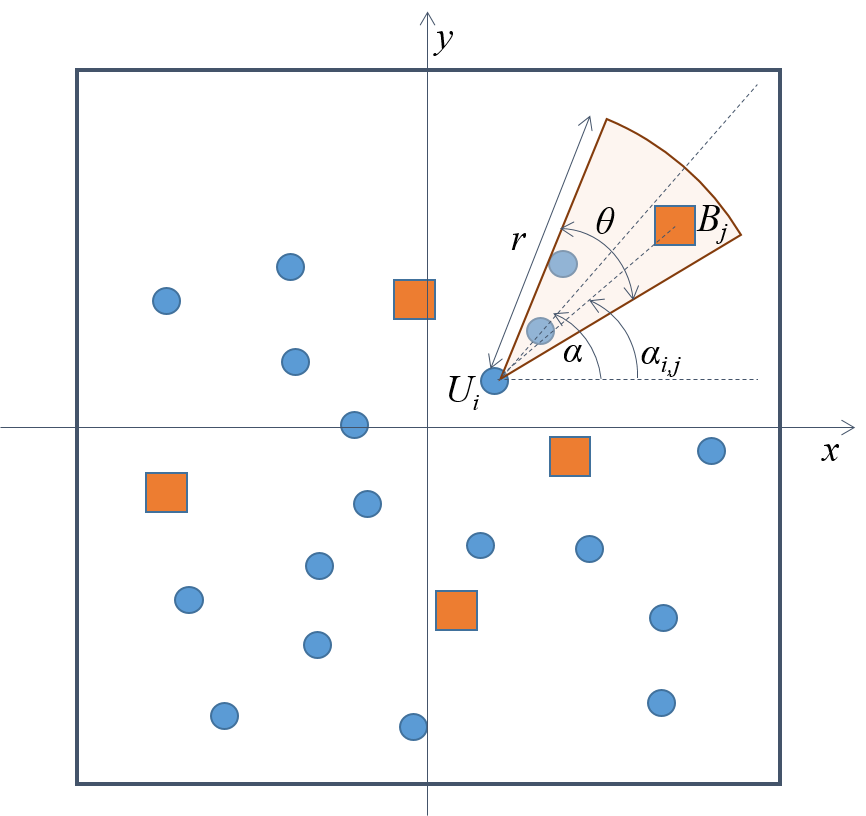}
\caption{User directional transmission model: User $U_i$ transmits a signal of range $r$ and beam-width $\Theta$ in a randomly oriented direction $\alpha$.}
\label{fig: System Model}
\end{figure}

In contrast with our previous work on SA-MBS \cite{SA-MBS}, in this paper we assume that UEs use \emph{directional antennas}, e.g., by exploiting beamforming techniques, to direct their transmission beams \cite{mmW-BF}. We assume a simple randomized beamforming model in which UEs choose the main lobe direction $\alpha$ uniformly at random from the interval $[0,2\pi)$, as presented in Figure \ref{fig: System Model} \cite{RBF}. The simple beamforming model avoids inefficiency due to beam-alignment procedures, while relying on assumption that the density of infrastructure $\lambda_{BS}$ (e.g., small cells) is very large. The main lobe angular width $\theta$ and the UE signal range $r$ is assumed equal for all UEs. In other words, we assume homogeneous model where each UE transmits using the same power, and simplify the signal propagation model by considering only path loss (shadowing and fading effects are neglected).

For the above model, a user transmission has a simple geometric interpretation where UE signals cover circular section areas centred at the UE location $x$ with the radius $r$, angular width $\theta$ and randomly selected angular offset $\alpha$, as illustrated in Figure \ref{fig: System Model}. We denote the surface covered by this circular section as $\mathcal{A}_{CS}(x,r,\alpha,\theta)$. A necessary condition for the UE to be collected is that $\mathcal{A}_{CS}(x,r,\alpha,\theta)$ covers at least one BS. We formalize the notion of an UE covering a BS as follows: the user $U_{i}$ \emph{covers} the base station $B_{j}$ if and only if the corresponding \emph{coverage indicator} $C_{ij}=1$:
\begin{equation}
	C_{ij}= \twopartdef { 1 } {(d_{ij} \leq r)\ \wedge (\alpha - \frac{\theta}{2} \leq \alpha_{ij} \leq \alpha + \frac{\theta}{2})} {0} {},
\end{equation}
where $d_{ij}$ is the Euclidean distance between $U_{i}$ and $B_{j}$, and $\alpha_{ij}$ is the angle of the line $(U_{i},B_{j})$ relative to the reference direction.

\begin{figure*}
\centering
\includegraphics[width=6.4in]{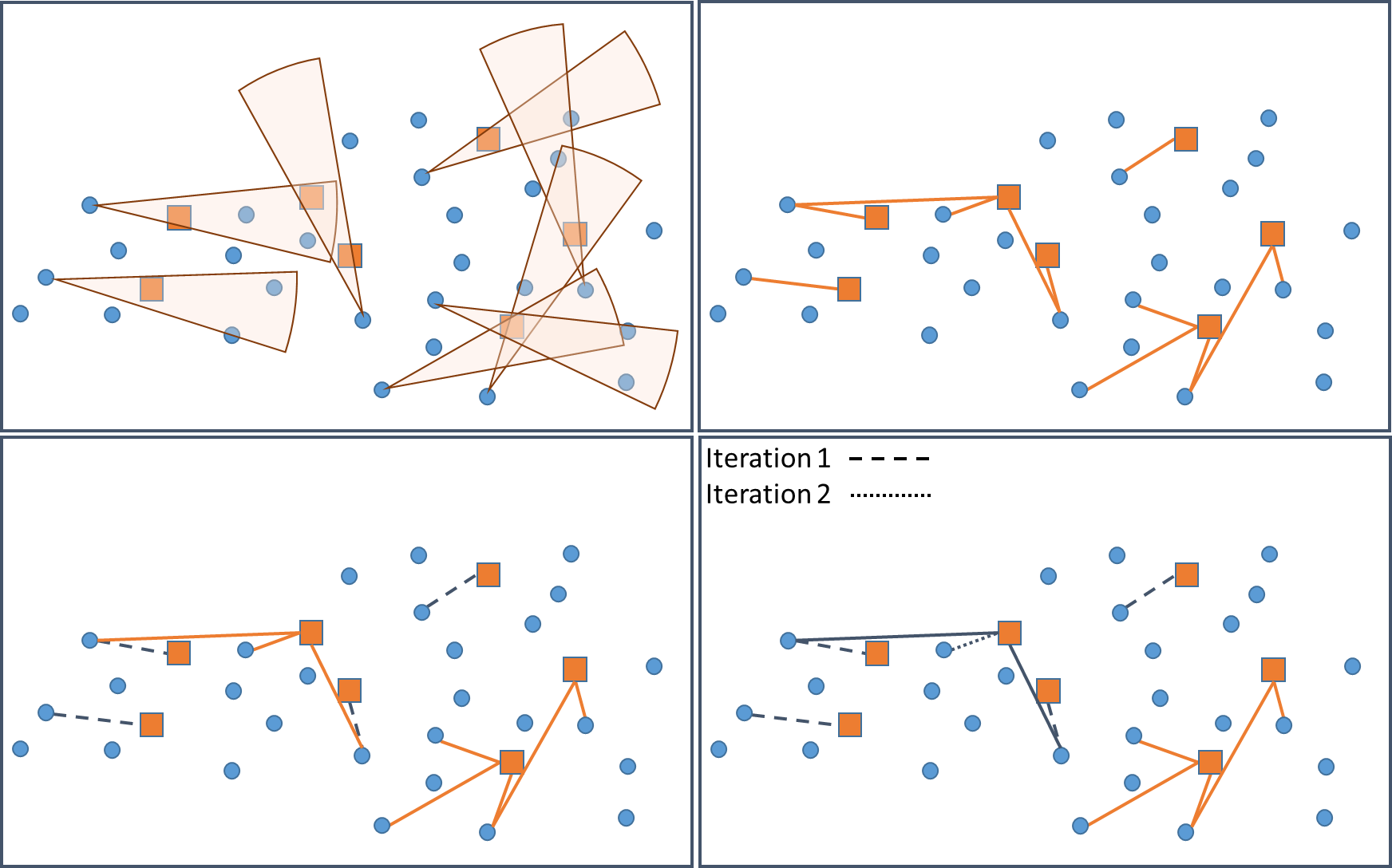}
\caption{Upper-left subfigure: Randomly directed UE transmissions in a given time slot; Upper-right subfigure: Resulting network connectivity graph; Lower-left subfigure: Non-cooperative SA-MBS decoding example - only UEs connected via dashed edges will be collected; Lower-right subfigure: Cooperative SIC-based SA-MBS decoding example - the first iteration is identical to non-cooperative decoding; the SIC phase removes solid lines; the second iteration decodes UEs connected via dotted lines (note that the set of four UEs in the lower-right corner cannot be decoded as it forms a \emph{stopping set} \cite{MCT}).}
\label{Fig_1}
\end{figure*}

We find the motivation for the randomized beamforming in its simplicity and the fact that it could provide a lower bound performance limit for more sophisticated beamforming techniques. To some extent, randomized beamforming simplifies our analysis, i.e., the specific performance approximations presented in the next section.

\subsection{Network Connectivity Graph}\label{NCG}

The above introduced coverage indicator $C_{ij}$ allows us to define the (random) \textit{Network Connectivity Graph} $\mathcal{C}=(V,E)$ as a bipartite graph of nodes $V=\{\mathcal{U},\mathcal{B}\}$, where $\mathcal{U}=\{U_1,U_2,\ldots,U_{N_{UE}}\}$ is the set of UEs and $\mathcal{B}=\{B_1,B_2,\ldots,B_{N_{BS}}\}$ is the set of BSs, and edges $E \subseteq \mathcal{U}\times\mathcal{B}$ such that $(U_i,B_j)\in E \iff C_{ij}=1$ (see Figure \ref{Fig_1}, upper subfigures). 
Note that the resulting graph represents a \emph{random bipartite geometric graph} and has in particular a random number of nodes~$N_{BS}+N_{UE}$.

For any user $U_{i}$, its degree $deg(U_{i})$ (number of adjacent BSs) in graph $\mathcal C$ is a Poisson random variable with mean $\frac{1}{2}\lambda_{BS}\,r^2\theta$, which we denote by $\mathcal{P}(\frac{1}{2}\lambda_{BS}\,r^2\theta)$. 
Similarly, for any BS $B_j$, its degree $deg(B_{j})$ (number of adjacent users) has the distribution $\mathcal{P}(\frac{1}{2}\lambda_{UE}\,r^2\theta)$ (we note that for both, UE and BS degree distributions, we ignore the boundary effects.)  Denote by $\Lambda_{d}=\mathbb{P}(deg(U_{i})=d)$, and by 
 $\Omega_{d}=\mathbb{P}(deg(B_{j})=d),$ $d=0,1,2,...$ %
%
For future reference, we also introduce the corresponding \emph{degree distribution polynomials} $\Lambda(x)=\sum_{d=0}^{\infty}\Lambda_d x^d$ and $\Omega(x)=\sum_{d=0}^{\infty}\Omega_d x^d$.
Finally, we note that the network connectivity graph $\mathcal{C}$ is a useful tool not only to visualise, but to also interpret and analyse the SA-MBS decoding process, as described next.

\section{Decoding Algorithms and Throughput Analysis}

In this section, we present two decoding approaches for SA-MBS system: i) non-cooperative and ii) cooperative decoding. Providing exact closed-form expressions for the system throughput in both scenarios seems to be a formidable task; however, we approximate the throughput with a simple but insightful expressions.

\subsection{Base Stations Decoding Model}

\textbf{Non-Cooperative SA-MBS Decoding:} in this case, we assume all BSs apply classical Slotted ALOHA decoding algorithm independently of each other. In other words, in any time slot, a BS will collect a UE's packet if and only if that UE is the only one that covers the BS (``singleton''). In contrast, if BS detects empty TS (no UE cover the BS) or TS is occupied by two or more UE transmissions, the TS, at that BS, is wasted. Non-cooperative SA-MBS decoding proceeds on a "slot-by-slot" basis, where TSs are independent among each other. In terms of the network connectivity graph, the algorithm allows simple interpretation: only BSs with degree equal to one are able to collect a corresponding UE (Figure \ref{Fig_1}, lower-left subfigure).

\textbf{Cooperative SA-MBS Decoding:} in this scenario, we assume all signals collected at BSs are forwarded to the central processing location. Motivation for this assumption comes from the so-called Cloud-RAN (C-RAN) architecture, where BSs serve only as RF front-ends while the baseband processing is done centrally. For simplicity, we assume all UE signals are synchronized to the TS boundaries at all BSs (i.e., the distance differences can be neglected due to high density of both UEs and BSs), and that BSs know and share with the centralized processing location the channel state information of the UEs in their vicinity. Centralized cooperative decoding algorithm applies SA with Successive Interference Cancellation (SIC) across all the signals received at different BSs \cite{CRDSA}. In short, if UE's transmission is decoded as a singleton at any BS, its signal can be subtracted from collisions at all other BSs where a given UE signal is found in collision with other UE signals. In terms of graphical interpretation, the signal recovery using cooperative SA-MBS algorithm on the network connectivity graph is equivalent to the iterative erasure decoding of LDPC codes \cite{IRSA}\cite{SA-MBS} performed on random bipartite geometric graphs (Figure \ref{Fig_1}, lower-right subfigure).

\subsection{Throughput Analysis}

We consider the SA-MBS system with average normalized load $G=\frac{\lambda_{UE}}{\lambda_{BS}}$. 
The average normalized system throughput is defined as the expected number of collected users per BS, per TS: 
 \[
T=\frac{1}{\lambda_{BS}}\mathbb{E}\left[ N_{UE,\,coll.} \right],
 \]
where $N_{UE,coll.}$ is the total number of collected users. Here, expectation is taken over the number of users, number of BSs, as well as over their random placements. 
 In the following, we consider the average normalized system throughput for the SA-MBS system with non-cooperative decoding.
 
\textbf{Non-Cooperative SA-MBS Decoding -- Throughput Upper Bound:} Denote by $N_{BS,1}$ the number of BSs which have degree equal to~$1$, i.e., which are in the transmission range of exactly one active user. Note that, for the random variables $N_{UE,coll.}$ and $N_{BS,1}$, it holds that $N_{UE,coll.} \leq N_{BS,1}$. Indeed, for each collected user~$U_i$, there exists at least one BS which collected it, and this BS necessarily has degree~$1$. Note also that there might be more than one BS which collects $U_i$; and all such BSs have degree one and are adjacent to $U_i$. Moreover, the set of the BSs which collect $U_i$ and the set of the BSs which collect $U_j$ are \emph{disjoint}, for any $i \neq j$. Now, having that $N_{UE,coll.} \leq N_{BS,1}$, we can upper bound the average normalized system throughput as follows:
 \begin{equation}
 \label{eqn-T-bnd}
 T \leq \frac{1}{\lambda_{BS}} \mathbb{E} \left[ N_{BS,1}\right].
 \end{equation}
 It remains to calculate $\mathbb{E} \left[ N_{BS,1}\right]$. We do so by conditioning on the number of BSs $N_{BS}$, i.e.:
 \begin{equation*}
 \mathbb{E} \left[ N_{BS,1}\right] = \sum_{b=0}^{\infty}
 \mathbb{E} \left[ N_{BS,1}\,|\,N_{BS}=b\,\right]\,\mathbb{P}(N_{BS}=b).
 \end{equation*} 

Next, it is clear that $\mathbb{E} \left[ N_{BS,1}\,|\,N_{BS}=b\,\right] = b\, \Omega_1,$ where we recall that $\Omega_1$ 
is the probability that a fixed BS has degree~$1$, and it equals $\mu\,e^{-\mu}$, with~$\mu = \frac{\lambda_{UE} r^2 \theta}{2}$. Now, using $\mathbb{P}(N_{BS}=b) = e^{-\lambda_{BS}} \frac{\lambda_{BS}^b}{b!}$, we have:
  \begin{eqnarray*}
  \mathbb{E} \left[ N_{BS,1}\right] &=& \sum_{b=0}^{\infty}
 b\,\mu\,e^{-\mu}\,\frac{e^{-\lambda_{BS}}\lambda_{BS}^b}{b!}\\
 &=& \lambda_{BS} \,\,\mu\,e^{-\mu} \sum_{b=1}^{\infty}
   \frac{e^{-\lambda_{BS}}\lambda_{BS}^{b-1}}{(b-1)!} 
   = \lambda_{BS} \,\,\mu\,e^{-\mu}
 .
 \end{eqnarray*}  
   Substituting the latter expression in~\eqref{eqn-T-bnd}, and using $\mu =\frac{\lambda_{UE}\,r^2 \,\theta}{2}$, 
   we finally obtain the following upper bound on the average normalized throughput:
   \begin{eqnarray}
   \label{eqn-t-bnd-2}
   T &\leq& \frac{\lambda_{UE}\,r^2 \,\theta}{2} \,\mathrm{exp} \left( - \frac{\lambda_{UE}\,r^2 \,\theta}{2}\right)\\
   &=& \frac{G\,\lambda_{BS}\,r^2 \,\theta}{2} \,\mathrm{exp} \left( - \frac{G\,\lambda_{BS}\,r^2 \,\theta}{2}\right). \nonumber
   \end{eqnarray}
   We note that the form of the above upper bound corresponds exactly to the well-known exact throughput expression for SA in single BS systems. As noted above, the throughput penalty for SA-MBS system that use non-cooperative decoding arises from the fact that some UEs will be collected by multiple BSs.  
  
   It is now interesting to investigate the bound~\eqref{eqn-t-bnd-2} as a function of system parameters $r$ and $\theta$, when $\lambda_{UE}$, $\lambda_{BS}$, and $G$ are fixed. Note that the bound is of the form $z\,e^{-z}$, with $z = \frac{\lambda_{UE}\,r^2 \,\theta}{2}$. The function $z \,e^{-z}$ has its maximum at~$z=1$, and hence the maximal throughput is achieved when $r$ and $\theta$ are such that $\frac{\lambda_{UE}\,r^2 \,\theta}{2} =1$. In words, the maximal throughput (bound) is achieved when $r$ and $\theta$ are such that, on average, each BS ``sees'' exactly one active user. Simulations further ahead will show that the optimal parameter estimates based on bound~\eqref{eqn-t-bnd-2} are close to the actual optimal system parameters.
   
\textbf{Cooperative SA-MBS Decoding - Throughput Upper Bound:} in this scenario, after all signals in a TS are collected and forwarded to the central processing location, the cooperative SIC-based decoding operates on the corresponding network connectivity graph $\mathcal{C}$. The graph contains on average $\overline{N}_{UE}=\lambda_{UE}$ user nodes and $\overline{N}_{BS}=\lambda_{BS}$ base station nodes, while edges follow UE-to-BS coverage indicators. The degree distributions of both UE and BS nodes can be approximated as Poisson distributions provided in subsection \ref{NCG}. 

We consider the asymptotic scenario when both $\lambda_{UE}$ and $\lambda_{BS} \rightarrow \infty$, while keeping constant the average system load $G=\frac{\lambda_{UE}}{\lambda_{BS}}$. We also fix to a constant value the average UE degree $\bar{d}_{UE}=\frac{1}{2}\lambda_{BS}r^2\theta$ and the average BS degree $\bar{d}_{BS}=\frac{1}{2}\lambda_{UE}r^2\theta$. In the asymptotic limit, the UE and BS degree distributions tend to Poisson distributions $\Lambda(x)=e^{-\bar{d}_{UE}(1-x)}=e^{-G\cdot\bar{d}_{BS}(1-x)}$ and $\Omega(x)=e^{-\bar{d}_{BS}(1-x)}$. In such a scenario, the asymptotic probability that a UE is collected after $l$ iterations of cooperative SA-MBS decoding, under assumption that the underlying graph does not contain cycles of length $2l$ or less, can be obtained by applying the density evolution analysis \cite{MCT}. 

Let $\mathbb{P}^{(l)}(U_{i,coll.})$ be the probability that a UE $U_i$ is collected after the $l$-th iteration of iterative SIC-based cooperative SA-MBS decoding. Observing the SIC-based decoding process as graph-peeling iterative erasure decoder running on the network connectivity graph, we denote by $q_l$ ($r_l$) the average probability that an edge in the graph incident to a user (base station) survives the $l$-th iteration. Then, by and-or-tree lemma \cite{and-or-tree}, we have:
\begin{eqnarray}
r_l&=&1-e^{-(1-G)\cdot\bar{d}_{BS}\cdot(1-q_{l-1})};\\
q_l&=&e^{-G\cdot\bar{d}_{BS}(1-r_l)};
\end{eqnarray}
where the recursion is initialized by $q_0=0$. The expected probability a user is collected after the $l$-th iteration of the iterative SIC-based decoder is equal $\mathbb{P}^{(l)}(U_{i,coll.})=1-q_l$.

\begin{figure}
\centering
\includegraphics[width=3.3in, height=2in]{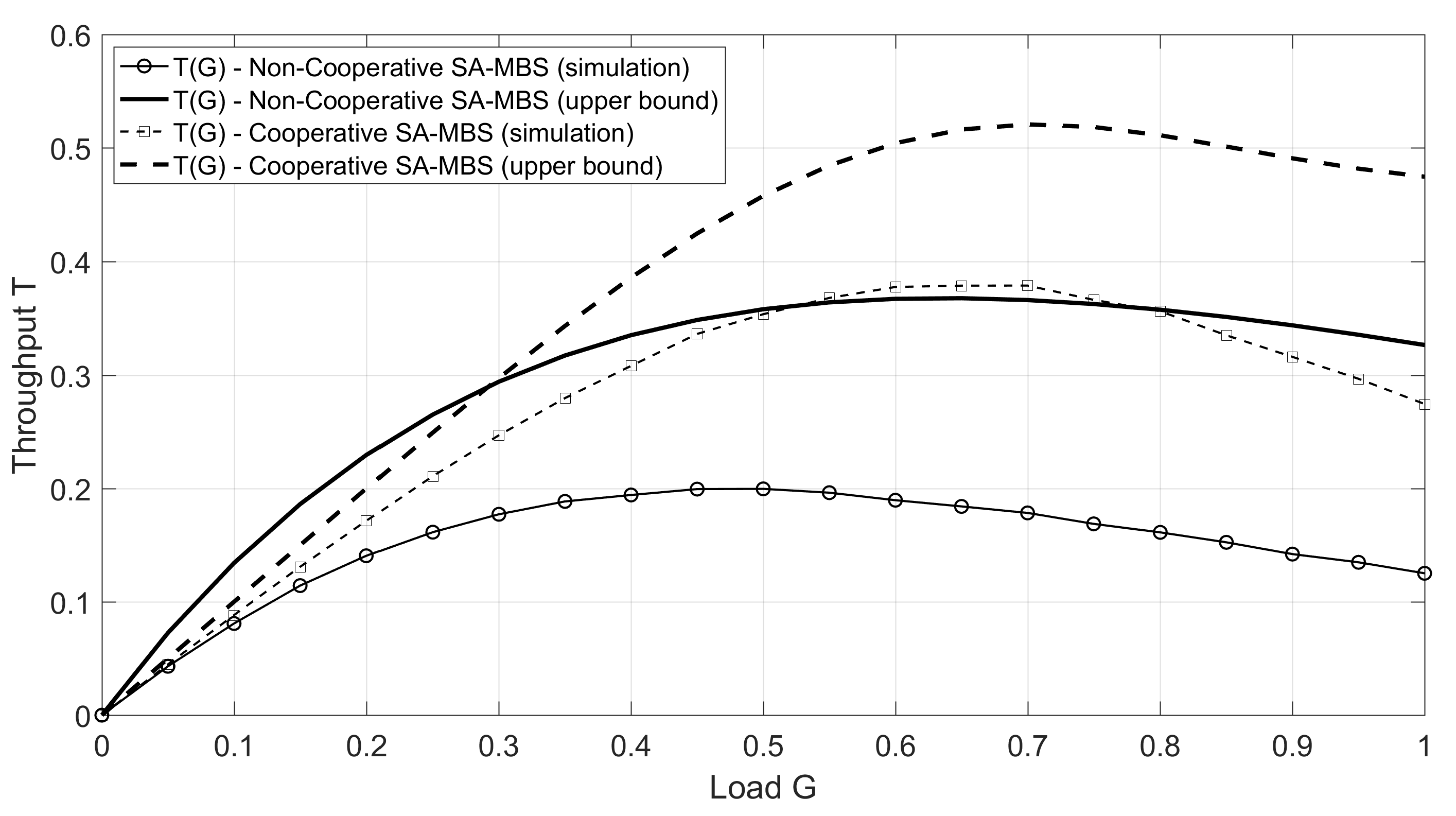}
\caption{Average system throughput $T=T(G)$ for cooperative and non-cooperative SA-MBS decoding (simulation results and upper bound).}
\label{Figure_3}
\end{figure}

We note that the output of the and-or-tree analysis overestimates $\mathbb{P}(U_{i,coll.})$ due to the fact that: i) in real-world systems, we deal with finite values of $\lambda_{UE}$ and $\lambda_{BS}$, and ii) unlike assumed in and-or-tree analysis, our network connectivity graph is random bipartite \emph{geometric} graph, and consequently, the probability of short cycles will not vanish asymptotically. Thus we adopt the output of and-or-tree analysis as an upper bound for $\mathbb{P}(U_{i,coll.})$ in case of cooperative SA-MBS decoding, as demonstrated numerically in the following section.   

\section{Numerical Results}

\begin{figure}
\centering
\includegraphics[width=3.3in, height=2in]{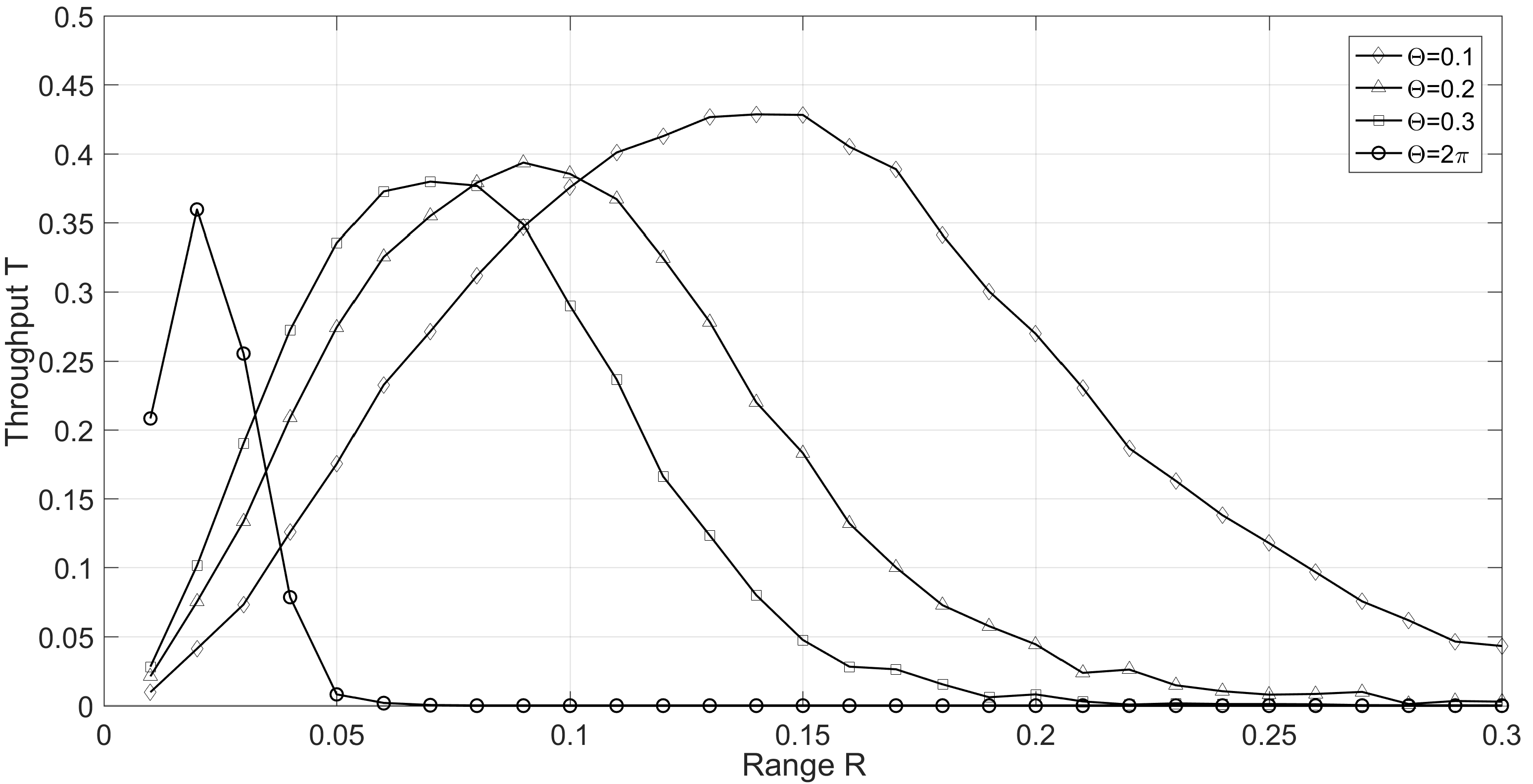}
\caption{Average system throughput $T=T(r)$ for cooperative SA-MBS for different beam-widths $\Theta$ and load $G=1$.}
\label{Figure_4}
\end{figure}

In this section, we present the average system throughput results obtained via simulation experiments and discuss the performance of SA-MBS with directional antennas as a function of user beam-width $\Theta$ and beam-range $r$. We compare the simulation results with not very tight but simple and informative throughput upper bounds presented in Section 3B. 

The simulation setup is described as follows. In each simulation experiment, we fix the average system load $G=\frac{\lambda_{UE}}{\lambda_{BS}}$. Then, we sample $N_{UE} = n$ and $N_{BS} = b$ from $\mathcal{P}(\lambda_{UE})$ and $\mathcal{P}(\lambda_{BS})$, i.e., the UEs and BSs are placed over a unit-square area according to PPPs with intensities $\lambda_{UE}$ and $\lambda_{BS}$ (the two PPPs being independent). For all UEs, we generate their corresponding circular section areas $\mathcal{A}_{CS}$ as described in Section 2C, and use them to define the corresponding network connectivity graph $\mathcal{C}$. We run both decoding algorithms on $\mathcal{C}$, as described in Section 3A, in order to obtain the number of collected users $N_{UE,coll.}=n_{coll.}$. For a fixed set of parameters, the simulation experiment is repeated $100$ times to obtain the estimate of the average number of collected users $\overline{N}_{UE,coll.}$. The average system throughput is estimated as $T=G\cdot\frac{\overline{N}_{UE,coll.}}{\overline{N}_{UE}}=\frac{\overline{N}_{UE,coll.}}{\lambda_{BS}}.$  

In Figure \ref{Figure_3}, we present the average system throughput, both simulated and upper bounds, for non-cooperative and cooperative SA-MBS decoding. We fixed the average BS density to $\lambda_{BS}=1000$ and varied the average UE density $\lambda_{UE}$ in order to evaluate $T=T(G)$ over the range $G=[0,1]$. The directional antenna beam parameters are fixed to $r=0.1$ and $\theta=\pi/10$. The figure shows that the average throughput of cooperative SA-MBS decoding significantly outperformes the non-cooperative case and reaches the maximum $T^{*}\sim0.38$ for $G=0.7$, compared to $T^{*}\sim0.2$ achieved by the non-cooperative decoding at $G=0.5$. We note that the derived upper bounds, although not tight, correctly follow the throughput behaviour and, as demonstrated next, could be used to estimate the optimal directional antenna parameter settings. 

In Figure \ref{Figure_4}, we present the simulated results for the system throughput $T$ in cooperative SA-MBS decoding scenario as a function of radius  $r$. We fix the average system load $G=1$ by fixing $\lambda_{UE}=\lambda_{BS}=1000$, and vary the UE beam-range $r=[0.01, 0.3]$, for the set of directional antenna beam-widths $\theta=\{0.1, 0.2, 0.3, 2\pi]$. Figure \ref{Figure_4} shows clear benefits of using directional antennas in a given scenario, where not only the maximum throughput slightly increases for smaller beam-widths $\Theta$, but also, the region of beam-ranges $r$ for which the throughput remains at high values is dramatically larger. 

\begin{figure}
\centering
\includegraphics[width=3.3in, height=2in]{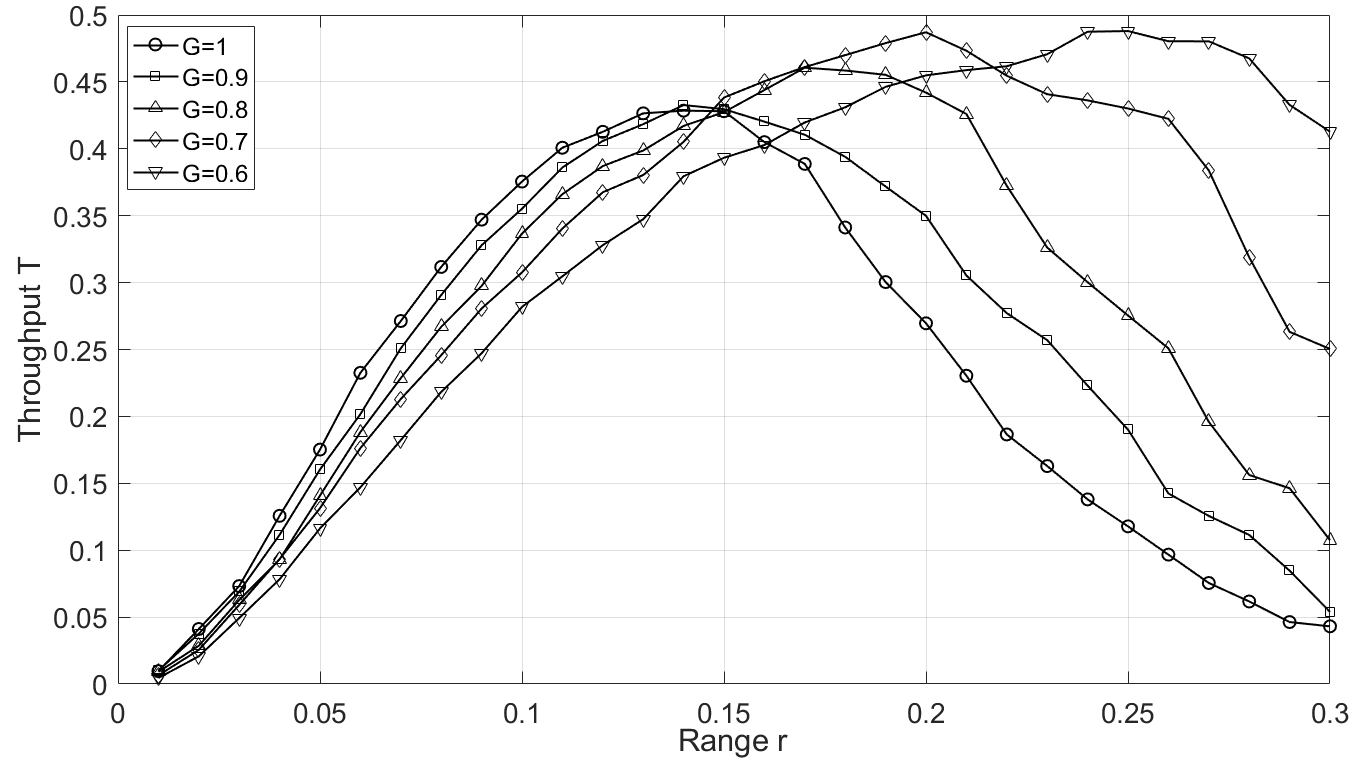}
\caption{Average system throughput $T=T(r)$ for cooperative SA-MBS for different loads $G$ and beam-width $\Theta=0.1$.}
\label{Figure_5}
\end{figure}

Figure \ref{Figure_5} investigates the system throughput $T$ as a function of radius $r$ for a fixed narrow beam-width $\Theta=0.1$. The average system load is varied between $G=[0.6,1]$ by varying $\lambda_{UE}$ while keeping $\lambda_{BS}=1000$. Figure \ref{Figure_5} shows how by decreasing the system load $G$, one can achieve high throughput for a wide range of beam-ranges $r$. The maximum throughput $T^{*}=0.48$ is achieved for $G=0.7$. 

\begin{figure}
\centering
\includegraphics[width=3.3in, height=2in]{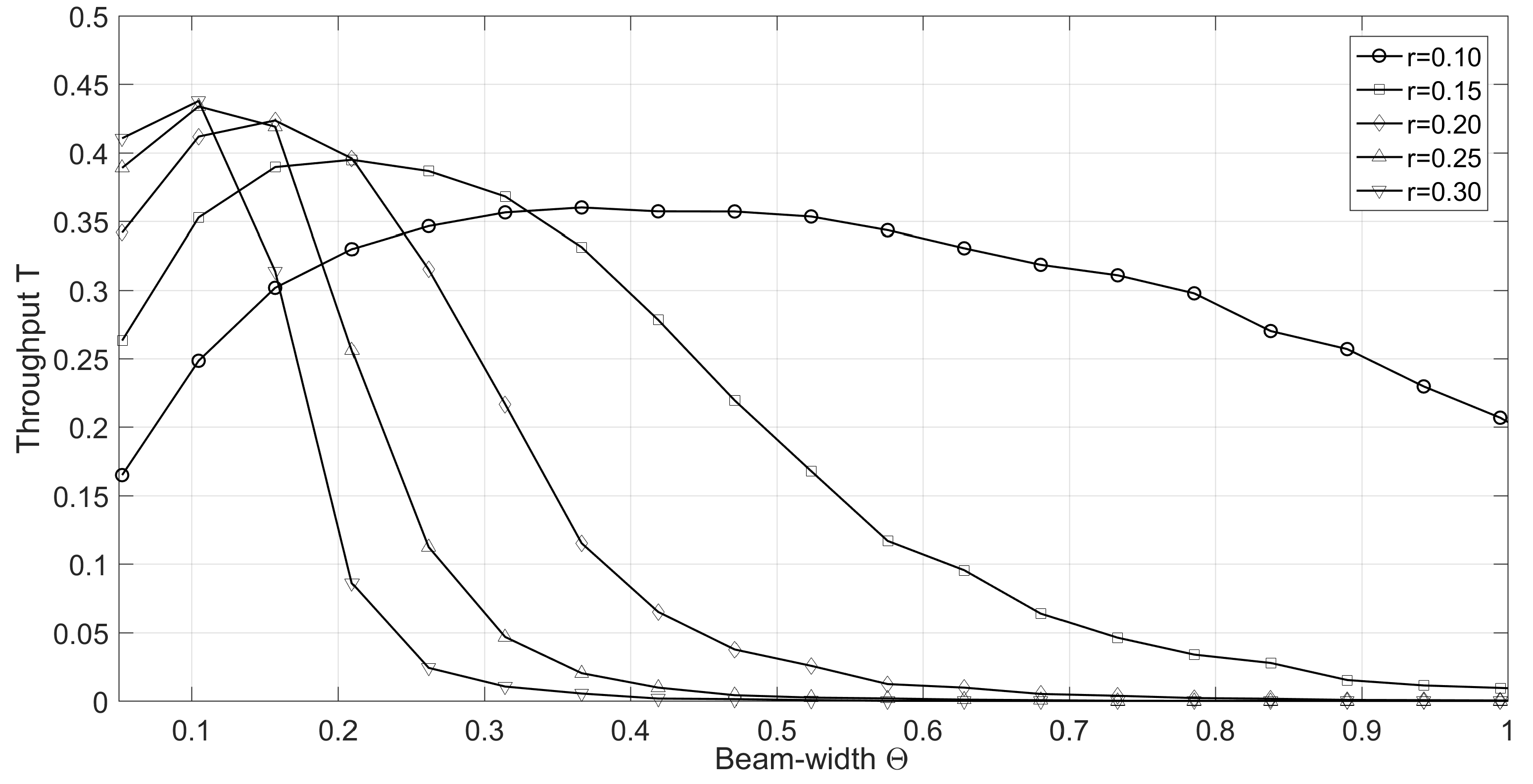}
\caption{Average system throughput $T=T(\Theta)$ for cooperative SA-MBS for different beam-ranges $r$ and load $G=0.5$.}
\label{Figure_6}
\end{figure}

Finally, in Figure \ref{Figure_6}, we consider the system throughput $T$ as a function of the beam-width $\Theta$ for a fixed load $G=0.5$ and a set of beam-ranges $r=\{0.1,0.15,0.2,0.25,0.3\}$. One can observe that the maximum throughput peaks at narrow beam-widths ($\Theta \approx 0.1$). As expected, for any beam-range $r$, there exists an optimum beam-width that maximizes the system throughput $T$. However, note that for smaller beam-range values $r$, the region of high throughput values stretches across considerably larger range of beam-widths $\Theta$.

\section{Conclusions}

In this paper, we investigated SA-MBS scenario considering users that employ directional antennas. The total system throughput is investigated for two decoding algorithms: the non-cooperative SA-MBS decoding where BSs independently apply traditional SA, and the cooperative SA-MBS decoding where signals received at BSs are centrally decoded using SIC-based decoding. Both scenarios are analyzed by evaluating the total system throughput using both simulation experiments and analytical throughput upper bounds. The obtained results demonstrate that the cooperative SA-MBS decoding can significantly outperform non-cooperative SA-MBS decoding. In addition, the obtained bounds could be used to provide guidelines on the selection of directional antenna parameters that, under a given system setting, maximize the total system throughput. As a future work, we intend to tighten analytical bounds for both non-cooperative and cooperative decoding scenario by considering the probabilities that a UE is collected by two or more BSs (non-cooperative) or by analyzing certain dominant stopping sets (cooperative). We will also extend the scenario where UEs employ directional antennas by additionally considering time-diversity, i.e., by exploiting Framed Slotted Aloha (FSA) and performing SIC-based decoding both spatially (across different BSs) and temporally (across different TSs).      

\section*{Acknowledgement}

The Research leading to these results has received funding from \textit{the EC Seventh Framework Programme} (\textbf{FP7-PEOPLE-2013-ITN}) under Grant Agreement No. 607774.

\begin{thebibliography}{1}

\bibitem{MTC}
T. Taleb, and A. Kunz, ``Machine type communications in 3GPP networks: potential, challenges, and solutions,'' \emph{IEEE Communications Magazine,} Vol. 50, No. 3, pp. 178-184, March 2012.

\bibitem{CRDSA}
E.~Casini, R.~{D}e Gaudenzi, and O.~del~rio Herrero, ``Contention resolution diversity slotted {ALOHA} ({CRDSA}): {A}n enhanced random access scheme for satellite access packet networks,'' \emph{IEEE Transactions on Wireless Communications}, Vol.~6, No.~4, pp. 1408--1419, April 2007.

\bibitem{IRSA}
G.~Liva, ``Graph-based analysis and optimization of contention resolution diversity slotted {ALOHA},'' \emph{IEEE Transactions on Communications}, Vol.~59, No.~2, pp. 477--487, February 2011.

\bibitem{SA-MBS}  
D. Jakovetic, D. Bajovic, D. Vukobratovic, V. Crnojevic: ``Cooperative Slotted ALOHA for Multi Base Station Systems,'' \emph{IEEE Transactions on Communications,} Vol. 63, No. 4, pp. 1443-1456, April 2015.

\bibitem{CSA}
E. Paolini, C. Stefanovic, G. Liva, P. Popovski, ``Coded Random Access: Applying Codes on Graphs to Design Random Access Protocols,'' \emph{IEEE Communications Magazine,} Vol. 53, No. 6, pp. 144-150, June 2015.

\bibitem{MBS}
A. Munari, F. Clazzer, G. Liva, ``Multi-Receiver Aloha - a Survey and New Results,'' in Proc. IEEE ICC Workshop on Massive Uncoordinated Access Protocols (MASSAP), London (UK), 8-12 June 2015. 

\bibitem{mmW-5G}
T. Rappaport, S. Sun, R. Mayzus, H. Zhao, Y. Azar, K. Wang, G.~N. Wong, J.~K. Schulz, M. Samimi, and F. Gutierrez, ``Millimeter wave mobile communications for 5G cellular: It will work!,'' \emph{IEEE Access}, Vol. 1, pp. 335-349, 2013.

\bibitem{mmW-BF}
W. Roh, J.-Y. Seol, J. Park, B. Lee, J. Lee, Y. Kim, J. Cho, K. Cheun, and F. Aryanfar, ``Millimeter-wave beamforming as an enabling technology for 5G cellular communications: theoretical feasibility and prototype results,'' \emph{IEEE Communications Magazine,} Vol. 52, No. 2, pp. 106-113, 2014.

\bibitem{RBF}
C. Bettstetter, C. Hartmann, and C. Moser, ``How does randomized beamforming improve the connectivity of ad hoc networks?,'' In Proc. IEEE ICC 2005, pp. 3380-3385, Seoul (Korea), May 2005.

\bibitem{5G-Disrupt}
F. Boccardi, R.~W. Heath, A. Lozano, T.~L. Marzetta, and P. Popovski, ``Five disruptive technology directions for 5G,'' \emph{IEEE Communications Magazine,} Vol. 52, No. 2, pp. 74-80, February 2014.

\bibitem{MCT}
T. Richardson, and R. Urbanke, ``Modern coding theory,'' \emph{Cambridge University Press,} 2008.

\bibitem{and-or-tree}
M. Luby, M. Mitzenmacher, and A. Shokrollahi, ``Analysis of random processes via and-or tree evaluation,'' in ACM SODA ’98, San Francisco, USA, 1998. 

\end{thebibliography}
\end{document}